\documentclass{elsart}
\usepackage{amsmath,amsfonts,amssymb}
\usepackage{graphics}
\usepackage{graphicx}

\begin{document}

\begin{frontmatter}















\title{Measuring quasiprobability distribution functions of the cavity field considering field and atomic decays}



\author{N. Yazdanpanah, M. K. Tavassoly}
\address{Atomic and Molecular Group, Faculty of Physics, Yazd University, Yazd, Iran}
\address{The Laboratory of Quantum Information Processing, Yazd University, Yazd, Iran}
\address{Photonic Research Group, Engineering Research Center, Yazd University, Yazd, Iran}
\author{R. Ju\'arez-Amaro}
\address{Universidad Tecnol\'ogica de la Mixteca, Apdo. Postal 71, Huajuapan de Le\'on, Oax., 69000 Mexico}
\address{Instituto Nacional de Astrof\'{i}sica, \'{O}ptica y Electr\'{o}nica \\ Calle Luis Enrique Erro No. 1, Sta. Ma. Tonantzintla, Pue. CP 72840, Mexico}
\author{ H. M. Moya-Cessa}
\address{Instituto Nacional de Astrof\'{i}sica, \'{O}ptica y Electr\'{o}nica \\ Calle Luis Enrique Erro No. 1, Sta. Ma. Tonantzintla, Pue. CP 72840, Mexico}

\begin{abstract}
We study the possibility of reconstructing the quantum state of light
in a cavity subject to dissipation. We pass atoms, also subject to decay,
through the cavity and surprisingly show that both decays allow the measurement of $s$-parametrized quasiprobability distributions. In fact, if we consider only atomic decay, we show that the Wigner function may be reconstructed. Because  these distributions contain whole information of the initial field state,  it is possible to recover information after both atomic and field decays occur. 
\end{abstract}


\end{frontmatter}
\section{Introduction}
The measurement of a quantum state is a central topic in
quantum optics and related fields \cite{ris,leo}. Several techniques have been developed in order to achieve such goal,  for instance tomographic reconstruction by unbalanced homodyning \cite{wal}, cascaded homodyning \cite{kis}, the
direct sampling of the density matrix of a signal mode in
multiport optical homodyne tomography \cite{zuk} and  reconstruction
via photocounting \cite{ban},  to
cite some. Proposals to measure
electromagnetic fields inside cavities \cite{lut,moy}  have also been given. Such
state reconstruction in cavities is usually achieved through a
finite set of selective measurements of atomic states \cite{lut}
that make it possible to construct different quasiprobability distribution
functions.

However, in real experiments, dissipative processes that have destructive effects may occur. Schemes that treat
dissipative cavities have been  proposed \cite{moy,moy2}. They involve
physical processes that allow the storage of information about
quantum coherences of the initial state in the diagonal elements
of the density matrix of a transformed state.

The relation between losses and
$s$-parametrized quasiprobability distributions has already been
pointed out by Leonhardt and Paul \cite{paul} and problems with the reconstruction of
the Wigner function have been analyzed in \cite{paul2}. Methods to
reconstruct quasiprobability distribution functions in cavity QED are usually based on
the expression \cite{Knight}
\begin{equation}
F(\alpha,s)=
\frac{1}{\pi(1-s)}\sum_{k=0}^{\infty}\left(\frac{s+1}{s-1}\right)^k
\langle \alpha,k|\hat \rho|\alpha,k\rangle,
\end{equation}%
with $s$ as the  parameter of quasiprobability function that indicates
 the relevant distribution ($s=-1$ Husimi \cite{Husimi},
$s=0$ Wigner \cite{Wigner} and $s=1$ Glauber-Sudarshan
\cite{Glauber,Sudarshan} distribution functions), $\rho$ the
density matrix and the states $|\alpha,k\rangle$ are the so-called
displaced number states \cite{Oliveira}.

Leibfried {\it et al.} \cite{Wineland} and Bertet {\it et al.}  \cite{Haroche}
used the above expression to measure the Wigner function ($s=0$
case) of the quantized motion of an ion and the quantized cavity
field, respectively. It is possible to reconstruct a quasiprobability distribution function  from
the above equation since  there is a direct recipe. Let
us write equation (1) as
\begin{equation} \label{quasi}
F(\alpha,s)=
\frac{1}{\pi(1-s)}\sum_{k=0}^{\infty}\left(\frac{s+1}{s-1}\right)^k
\langle k|\hat D^{\dagger}( \alpha) \hat \rho \hat  D(\alpha)|k\rangle,
\end{equation}%
where $D(\alpha)$ is the Glauber displacement operator \cite{Glauber}. Note that, in
order to obtain a quasiprobability distribution function  we need
to  displace the system by an amplitude $\alpha$
and then measure the diagonal elements of the (displaced) density
matrix.

We now aim to study
the problem of  reconstruction of the cavity field as
studied in \cite{Amaro}, {however,  not only cavity decay but also atomic decay is allowed}  \cite{trang}. We then want to show that it is still possible
to recover whole information about the initial state through the reconstruction of $s$-parametrized quasiprobability distributions.

\section{Dispersive interaction between a two-level atom and a quantized field}
We consider the master equation for the dispersive interaction
between a two-level atom and a quantized field
\begin{eqnarray} \label{ME}
\frac{d\hat{\rho}}{dt}=-i\chi [\hat{n}\hat{\sigma}_{z},\hat{\rho}]+ \hat{\mathcal L}_{F}\hat{\rho}+
\hat{\mathcal L}_{A}\hat{\rho},
\end{eqnarray}
with $\chi$  as the dispersive interaction constant, $\hat{n}=\hat{a}^{\dagger}\hat{a}$ where $\hat{a}^{\dagger}$ and $\hat{a}$ are the creation and annihilation operators, respectively. $\hat{\sigma}_{z}$ is the Pauli-spin matrix related to the atomic inversion, and $\hat{\rho}$ is the density matrix of  the atom-field system.

In the above equation the cavity and atomic decay terms are given by
\begin{eqnarray}
 \hat{\mathcal L}_j\hat{\rho}=(\hat{J}_j+\hat{L}_j)\hat{\rho}, \qquad j=A,F
\end{eqnarray}
in which the subscripts "A" and "F" refer to the terms atom and field, respectively. The corresponding superoperators are defined as
\begin{eqnarray}
 \hat{J}_A\hat{\rho}=2\Gamma \hat{\sigma}_-\hat{\rho}\hat{\sigma}_+, \qquad \hat{J}_F\hat{\rho}=2\gamma
 \hat{a}\hat{\rho} \hat{a}^{\dagger},
\end{eqnarray}
and \begin{eqnarray}
 \hat{L}_A\hat{\rho}=-\Gamma (\hat{\sigma}_+\hat{\sigma}_-\hat{\rho}+\hat{\rho}\hat{\sigma}_+\hat{\sigma}_-), \qquad \hat{L}_F\hat{\rho}=-\gamma (\hat{a}^{\dagger}\hat{a}\hat{\rho}+\hat{\rho} \hat{a}^{\dagger}\hat{a}).
\end{eqnarray}
We solve (\ref{ME}) by doing
\begin{eqnarray}
\hat{\rho}(t)=e^{(\hat{R}+\hat{L}_A+\hat{L}_F)t}e^{\int_0^t
e^{-(\hat{R}+\hat{L}_A+\hat{L}_F)t'}(\hat{J}_A+\hat{J}_F)e^{(\hat{R}+\hat{L}_A+\hat{L}_F)t'}dt'}\hat{\rho}(0),
\end{eqnarray}
where we have defined
\begin{eqnarray}
\hat{R}\rho=-i\chi \hat{\sigma}_z\hat{n}\hat{\rho}+i\chi\hat{\rho}\hat{\sigma}_z\hat{n}.
\end{eqnarray}
We note that
\begin{eqnarray}
&&e^{-(\hat{R}+\hat{L}_A+\hat{L}_F)t'}(\hat{J}_A+\hat{J}_F)e^{(\hat{R}+\hat{L}_A+\hat{L}_F)t'}\hat{\rho}
\\&=&\left(\hat{J}_Ae^{-(\hat{R}_F+2\Gamma)t}+\hat{J}_Fe^{-(\hat{R}_A+2\gamma)t}\right)\hat{\rho},
\end{eqnarray}
where the commutation relations
\begin{eqnarray}
[\hat{L}_A,\hat{J}_A]\hat{\rho}(t)&=&2\Gamma \hat{J}_A\hat{\rho}(t), \qquad [\hat{L}_F,\hat{J}_F]\hat{\rho}(t)=2\gamma \hat{J}_F\hat{\rho}(t),
\end{eqnarray}
and
\begin{eqnarray}
[\hat{R},\hat{J}_A]\hat{\rho}(t)=\hat{J}_A(\hat{R}_F \hat{\rho}(t)), \qquad [\hat{R},\hat{J}_F]\hat{\rho}(t)=\hat{J}_F(\hat{R}_A \hat{\rho}(t)),
\end{eqnarray}
have been used. In the above equations, we have defined the
superoperators $\hat{R}_F$ and $\hat{R}_A$ as
\begin{eqnarray}
\hat{R}_F\hat{\rho}=2i\chi \hat{n}\hat{\rho}-2i\chi \hat{\rho}\hat{n}, \qquad
\hat{R}_A\hat{\rho}= i\chi \hat{\sigma}_z\hat{\rho}-i\chi \hat{\rho}\hat{\sigma}_z,
\end{eqnarray}
which commute with all the other superoperators involved, such that
the solution of the evolved density matrix is written as
\begin{eqnarray}\label{11}
\hat{\rho}(t)=e^{(\hat{R}+\hat{L}_A+\hat{L}_F)t}e^{\hat{J}_A\frac{1-e^{-(\hat{R}_F+2\Gamma)t}}{\hat{R}_F+2\Gamma}}e^{\hat{J}_F\frac{1-e^{-(\hat{R}_A+2\gamma)t}}{\hat{R}_A+2\gamma}}\hat{\rho}(0).
\end{eqnarray}
One should note that, since $\hat{J}_A^2\hat{\rho}(t)=0$, the following simplification may be obtained
\begin{eqnarray}
e^{\hat{J}_A\frac{1-e^{-(\hat{R}_F+2\Gamma)t}}{\hat{R}_F+2\Gamma}}\hat{\rho}(t)=\left(1+\hat{J}_A\frac{1-e^{-(\hat{R}_F+2\Gamma)t}}{\hat{R}_F+2\Gamma}\right)\hat{\rho}(t),
\end{eqnarray}
and therefore,
\begin{eqnarray}\label{15.5}
\hat{\rho}(t)=\hat{\rho}_{1}(t)+\hat{\rho}_{2}(t),
\end{eqnarray}
where
\begin{eqnarray}\label{15}
\hat{\rho}_{1}(t)&=&e^{t(\hat{R}+\hat{L}_A+\hat{L}_F)} \ \ e^{\hat{J}_F\frac{1-e^{-(\hat{R}_A+2\gamma)t}}{\hat{R}_A+2\gamma}}\hat{\rho}(0),\nonumber\\\hat{\rho}_{2}(t)&=&e^{t(\hat{R}+\hat{L}_A+\hat{L}_F)} \ \ \ \hat{J}_A\frac{1-e^{-(\hat{R}_F+2\Gamma)t}}{\hat{R}_F+2\Gamma} \ \ e^{\hat{J}_F\frac{1-e^{-(\hat{R}_A+2\gamma)t}}{\hat{R}_A+2\gamma}}\hat{\rho}(0).
\end{eqnarray}
\subsection{Calculation of $\hat{\rho}_1(t)$}
In order to calculate the evolved density matrix (\ref{15.5}) it is needed to find both term on the right hand side of that equation.
For this, we consider the atom and field to be initially in arbitrary states such that the initial atom-field density operator reads as
\begin{eqnarray}\label{12}
\hat{\rho}(0)=\hat{\rho}_A(0)\hat{D}^{\dagger}(\alpha)\hat{\rho}_F(0)\hat{D}(\alpha),
\end{eqnarray}
in which $\hat{\rho}_A(0)=|\psi_A(0)\rangle\langle\psi_A(0)|$ with
\begin{eqnarray}
|\psi_A(0)\rangle=\sin\theta|e\rangle+\cos\theta|g\rangle,
\end{eqnarray}
where $\theta$ is an arbitrary angle and $|e\rangle$ ($|g\rangle$) is the excited (ground) state of the atom. We have considered an initially displaced (by an amplitude $\alpha$) arbitrary field. We rewrite equation (\ref{12}) with the help of the atomic operators as
\begin{eqnarray}\label{14}
\hat{\rho}(0)=\Big(\sin^{2}\theta \ \hat{\sigma}_{z}+\hat{\sigma}_{-}\hat{\sigma}_{+}+\sin(2\theta) \ \hat{\sigma}_{x}\Big)\hat{D}^{\dagger}(\alpha)\hat{\rho}_F(0)\hat{D}(\alpha),
\end{eqnarray}
where $\hat{\sigma}_{x}=(\hat{\sigma}_{+}+\hat{\sigma}_{-})/{2}$.
By acting the superoperator $\hat{R}_A$ on the atomic operators one arrives at
\begin{equation}
\hat{R}_A\hat{\sigma}_{z}=0, \qquad   \hat{R}_A\hat{\sigma}_{-}\hat{\sigma}_{+}=0,  \qquad  \hat{R}_A\hat{\sigma}_{x}=-2\chi\hat{\sigma}_{y}, \qquad \hat{R}_A\hat{\sigma}_{y}=2\chi\hat{\sigma}_{x}.
\end{equation}
In order to find the evolved density matrix, we need to calculate $\hat{\rho}_{1}(t)$ in (\ref{15}), so we have
\begin{eqnarray}\label{17}
e^{\hat{J}_F\frac{1-e^{-(\hat{R}_A+2\gamma)t}}{\hat{R}_A+2\gamma}}\hat{\rho}(0)&=&\sum_{m=0}^{\infty}\frac{1}{m!}\hat{J}_F^{m} \ \hat{D}^{\dagger}(\alpha)\hat{\rho}_F(0)\hat{D}(\alpha)\Big(\frac{1-e^{-(\hat{R}_A+2\gamma)t}}{\hat{R}_A+2\gamma}\Big)^{m} \ \hat{\rho}_A(0)\nonumber\\&=&\sum_{m=0}^{\infty}\frac{1}{m!} \ (2\gamma)^{m} \ \hat{a}^{m} \hat{D}^{\dagger}(\alpha)\hat{\rho}_F(0)\hat{D}(\alpha) \hat{a}^{\dag m}\ \Big(\int_{0}^{t}dt^{'}e^{-(\hat{R}_A+2\gamma)t^{'}}\Big)^{m} \ \hat{\rho}_A(0),\nonumber\\
\end{eqnarray}
with
\begin{eqnarray} \label{18}
\int_{0}^{t}dt^{'}e^{-(\hat{R}_A+2\gamma)t^{'}} \ \hat{\rho}_A(0)&=&\int_{0}^{t}dt^{'}e^{-2\gamma t^{'}}e^{-\hat{R}_A t^{'}} \ \hat{\rho}_A(0)\nonumber\\&=&\int_{0}^{t}dt^{'}e^{-2\gamma t^{'}}e^{-i\chi\hat{\sigma}_{z} t^{'}} \ \hat{\rho}_A(0)  \ e^{i\chi\hat{\sigma}_{z} t^{'}}\nonumber\\&=&\int_{0}^{t}dt^{'}e^{-2\gamma t^{'}}\Big\{\sin^{2}\theta \ \hat{\sigma}_{z}+\hat{\sigma}_{-}\hat{\sigma}_{+}\nonumber\\&+&\sin(2\theta) \ \Big(\hat{\sigma}_{x}\cos(2\chi t^{'})+\hat{\sigma}_{y}\sin(2\chi t^{'})\Big)\Big\}\nonumber\\&=&\frac{1-e^{-2\gamma t}}{2\gamma}\Big(\sin^{2}\theta \ \hat{\sigma}_{z}+\hat{\sigma}_{-}\hat{\sigma}_{+}\Big)\nonumber\\&+&\sin(2\theta)\Big\{\hat{\sigma}_{x}{\mathcal C}(\chi,\gamma,t)+\hat{\sigma}_{y}{\mathcal S}(\chi,\gamma,t)\Big\},
\end{eqnarray}
where we have defined the following abbreviations
\begin{eqnarray}
{\mathcal C}(\chi,\gamma,t)&=&\int_{0}^{t}dt^{'}e^{-2\gamma t^{'}}\cos(2\chi t^{'})=\frac{-2\gamma\cos(2\chi t)+2\chi\sin(2\chi t)}{4\chi^2+4\gamma^2}e^{-2\gamma t}\nonumber\\&+&\frac{2\gamma}{4\chi^2+4\gamma^2},\nonumber\\
{\mathcal S}(\chi,\gamma,t)&=&\int_{0}^{t}dt^{'}e^{-2\gamma t^{'}}\sin(2\chi t^{'})=\frac{-2\gamma\sin(2\chi t)-2\chi\cos(2\chi t)}{4\chi^2+4\gamma^2} e^{-2\gamma t}\nonumber\\&+&\frac{2\chi}{4\chi^2+4\gamma^2}.
\end{eqnarray}
In obtaining equation (\ref{18}) we have used the following relations
\begin{eqnarray}
e^{-i\chi\hat{\sigma}_{z} t^{'}} \ \hat{\sigma}_{x}  \ e^{i\chi\hat{\sigma}_{z} t^{'}}&=& \hat{\sigma}_{x}\cos(2\chi t^{'})+\hat{\sigma}_{y}\sin(2\chi t^{'}),\nonumber\\e^{-i\chi\hat{\sigma}_{z} t^{'}} \ \hat{\sigma}_{y}  \ e^{i\chi\hat{\sigma}_{z} t^{'}}&=& \hat{\sigma}_{y}\cos(2\chi t^{'})-\hat{\sigma}_{x}\sin(2\chi t^{'}).
\end{eqnarray}
By using $\hat{\sigma}_{x}=(\hat{\sigma}_{+}+\hat{\sigma}_{-})/2$ and $\hat{\sigma}_{y}=(\hat{\sigma}_{+}-\hat{\sigma}_{-})/2i$ and defining $\zeta={\mathcal C}(\chi,\gamma,t)+i{\mathcal S}(\chi,\gamma,t)$ one can deduce
\begin{eqnarray}\label{21}
\Big(\int_{0}^{t}dt^{'}e^{-(\hat{R}_A+2\gamma)t^{'}}\Big)^{m} \ \hat{\rho}_A(0)&=&\Big(\frac{1-e^{-2\gamma t}}{2\gamma}\Big)^{m}\Big(\sin^{2}\theta \ \hat{\sigma}_{z}+\hat{\sigma}_{-}\hat{\sigma}_{+}\Big)\nonumber\\&+&\frac{1}{2}\sin(2\theta)\Big\{\hat{\sigma}_{+}\zeta^{\star m}+\hat{\sigma}_{-} \zeta^{m}\Big\}.
\end{eqnarray}
Utilizing (\ref{17}) and (\ref{21}) results in the expression
\begin{eqnarray}\label{23}
e^{\hat{J}_F\frac{1-e^{-(\hat{R}_A+2\gamma)t}}{\hat{R}_A+2\gamma}}\hat{\rho}(0)&=&\sum_{m=0}^{\infty}\frac{1}{m!} \ (2\gamma)^{m} \ \hat{a}^{m}\hat{D}^{\dagger}(\alpha)\hat{\rho}_F(0)\hat{D}(\alpha) \hat{a}^{\dag m}\nonumber\\&\times& \Bigg\{\Big(\frac{1-e^{-2\gamma t}}{2\gamma}\Big)^{m}\Big(\sin^{2}\theta \ \hat{\sigma}_{z}+\hat{\sigma}_{-}\hat{\sigma}_{+}\Big)\nonumber\\&+&\frac{1}{2}\sin(2\theta)\Big(\hat{\sigma}_{+}\zeta^{\star m}+\hat{\sigma}_{-} \zeta^{m}\Big)\Bigg\},\nonumber\\
\end{eqnarray}
we finally arrive at the end of our first task, namely the determination of $\hat{\rho}_1$
\begin{eqnarray}\label{25}
\hat{\rho}_{1}(t)&=&e^{-(\Gamma\hat{\sigma}_{+}\hat{\sigma}_{-}+\gamma \hat{n}+i\chi\hat{\sigma}_{z}\hat{n})t} \ \ \sum_{m=0}^{\infty}\frac{1}{m!} \ (2\gamma)^{m} \ \hat{a}^{m}\hat{D}^{\dagger}(\alpha)\hat{\rho}_F(0)\hat{D}(\alpha) \hat{a}^{\dag m} \nonumber\\&\times& \Bigg\{\Big(\frac{1-e^{-2\gamma t}}{2\gamma}\Big)^{m}\Big(\sin^{2}\theta \ \hat{\sigma}_{z}+\hat{\sigma}_{-}\hat{\sigma}_{+}\Big)\nonumber\\&+&\frac{1}{2}\sin(2\theta)\Big(\hat{\sigma}_{+}\zeta^{\star m}+\hat{\sigma}_{-} \zeta^{m}\Big)\Bigg\} \ \ e^{-(\Gamma\hat{\sigma}_{+}\hat{\sigma}_{-}+\gamma \hat{n}-i\chi\hat{\sigma}_{z}\hat{n})t}.
\end{eqnarray}
\subsubsection{Calculation of $\hat{\rho}_{2}(t)$}
Now we pay attention to the  calculation of $\hat{\rho}_{2}(t)$. First we note to the following  facts
\begin{eqnarray}
\hat{J}_A\hat{\sigma}_{\pm}&=&\hat{J}_A\hat{\sigma}_{-}\hat{\sigma}_{+}=0,\nonumber\\
\hat{J}_A\hat{\sigma}_{z}&=&2\Gamma\hat{\sigma}_{-}\hat{\sigma}_{+}=2\Gamma|g\rangle\langle g|.
\end{eqnarray}
Therefore, we have
\begin{eqnarray}
 \hat{J}_A \ \ e^{\hat{J}_F\frac{1-e^{-(\hat{R}_A+2\gamma)t}}{\hat{R}_A+2\gamma}}\hat{\rho}(0)=2\Gamma \sin^{2}\theta \sum_{m=0}^{\infty}\frac{(1-e^{-2\gamma t})^{m}}{m!} \ \hat{a}^{m} \hat{D}^{\dagger}(\alpha)\hat{\rho}_F(0)\hat{D}(\alpha) \hat{a}^{\dag m} \   |g\rangle\langle g|.\nonumber\\
\end{eqnarray}
Also, for each operator $\hat{x}$ one can obtain
\begin{eqnarray}
\frac{1-e^{-\hat{x}t}}{\hat{x}}=\sum_{l=0}^{\infty}(-\hat{x})^{l}\frac{t^{l+1}}{(l+1)!},
\end{eqnarray}
so that,
\begin{eqnarray}
\frac{1-e^{-(\hat{R}_F+2\Gamma)t}}{\hat{R}_F+2\Gamma}=\sum_{l=0}^{\infty}\frac{t^{l+1}}{(l+1)!}(-1)^{l}\sum_{k=0}^{l}\frac{l!}{k!(l-k)!}(2\Gamma)^{l-k}\hat{R}_F^{k},
\end{eqnarray}
where we have used the binomial theorem. For the superoperator $\hat{R}_F^{k}$
 we have
 \begin{eqnarray}
 \hat{R}_F^{k}\hat{\rho}_F=(2i\chi)^{k}\sum_{j=0}^{k}(-1)^{j}\frac{k!}{j!(k-j)!}\hat{n}^{k-j} \ \ \hat{\rho}_F \ \ \hat{n}^{j},
 \end{eqnarray}
and finally
\begin{eqnarray}\label{31}
\hat{\rho}_{2}(t)&=&2\Gamma\sin^{2}\theta\sum_{l=0}^{\infty}\frac{t^{l+1}}{(l+1)!}(-1)^{l}\sum_{k=0}^{l}\frac{l!}{k!(l-k)!}(2\Gamma)^{l-k}\nonumber\\&\times&\sum_{m=0}^{\infty}\frac{(1-e^{-2\gamma t})^{m}}{m!} \ (2i\chi)^{k}\sum_{j=0}^{k}(-1)^{j}\frac{k!}{j!(k-j)!}\nonumber\\&\times& e^{(i\chi-\gamma)\hat{n}t} \ \hat{n}^{k-j} \ \   \hat{a}^{m} \hat{D}^{\dagger}(\alpha)\hat{\rho}_F(0)\hat{D}(\alpha) \hat{a}^{\dag m} \   \ \hat{n}^{j} \ e^{(-i\chi-\gamma)\hat{n}t} \ |g\rangle\langle g|.
\end{eqnarray}

\section{Measuring quasiprobability distribution functions}

At this stage, to measure the quasiprobability   distributions,   we use the method that has been  proposed in Ref. \cite{Amaro}. To achieve this purpose,
one should measure the atomic polarization $\langle\hat{\sigma}_{x}\rangle$ as follows
\begin{eqnarray}\label{32}
\langle\hat{\sigma}_{x}\rangle&=&Tr\Big(\hat{\rho}(t)\hat{\sigma}_{x}\Big)=Tr\Big(\hat{\rho}_{1}(t)\hat{\sigma}_{x}\Big)\nonumber\\&=&\sum_{n=0}^{\infty}\langle e,n|\hat{\rho}_{1}(t)\hat{\sigma}_{x}|e,n\rangle+\langle g,n|\hat{\rho}_{1}(t)\hat{\sigma}_{x}|g,n\rangle\nonumber\\&=&\frac{1}{4}\sin(2\theta)e^{-\Gamma t}\sum_{n=0}^{\infty}e^{-2\eta n t}\sum_{m=0}^{\infty}\frac{(2\gamma\zeta^{\star})^m}{m!}\langle n| \hat{a}^{m} \hat{D}^{\dagger}(\alpha)\hat{\rho}_F(0)\hat{D}(\alpha) \hat{a}^{\dag m}  |n\rangle+c.c.\nonumber\\
\end{eqnarray}
where $\eta=\gamma+i\chi$ and $Tr\Big(\hat{\rho}_{2}(t)\hat{\sigma}_{x}\Big)=0$.
In the above relation, $\hat{\rho}_{1}(t)$
 and $\hat{\rho}_{2}(t)$ have been defined in  (\ref{25}) and (\ref{31}), respectively.
Equation (\ref{32}) can be simplified as
\begin{eqnarray}
\langle\hat{\sigma}_{x}\rangle=\frac{1}{4}\sin(2\theta)e^{-\Gamma t}\sum_{n=0}^{\infty}e^{-2\eta n t}\sum_{m=0}^{\infty}\frac{(2\gamma\zeta^{\star})^m}{m!}\frac{(n+m)!}{n!}\langle n+m| \hat{D}^{\dagger}(\alpha)\hat{\rho}_F(0)\hat{D}(\alpha)   |n+m\rangle+c.c.\nonumber\\
\end{eqnarray}
Since
\begin{eqnarray}
2\gamma\zeta^{\star}=\frac{\gamma}{\eta}(1-e^{-2\eta t}),
\end{eqnarray}
we have
\begin{eqnarray}
\langle\hat{\sigma}_{x}\rangle=\frac{1}{4}\sin(2\theta)e^{-\Gamma t}\sum_{n=0}^{\infty}e^{-2\eta n t}\sum_{m=0}^{\infty}\Big(\frac{\gamma}{\eta}(1-e^{-2\eta t})\Big)^{m}\frac{(n+m)!}{m!n!}\langle n+m|  \hat{\rho}_F(0)   |n+m\rangle+c.c.\nonumber\\
\end{eqnarray}
By changing the summation index in the second sum of the
above equation with $n+m=k$, one may obtain
\begin{eqnarray}\label{36}
\langle\hat{\sigma}_{x}\rangle=\frac{1}{4}\sin(2\theta)e^{-\Gamma t}\sum_{n=0}^{\infty}e^{-2\eta n t}\sum_{k=n}^{\infty}\Big(\frac{\gamma}{\eta}(1-e^{-2\eta t})\Big)^{k-n}\frac{k!}{(k-n)!n!}\langle k| \hat{D}^{\dagger}(\alpha)\hat{\rho}_F(0)\hat{D}(\alpha)  |k\rangle+c.c.\nonumber\\
\end{eqnarray}
It is noticeable that equation (\ref{36}) can be reduced to equation (20) of Ref. \cite{Amaro}
 via choosing $\Gamma=0$ (without considering the atomic decay) and $\theta=\frac{\pi}{4}$ (especial case) as is expected.

We now start the second sum from $k=0$ (as we would add only zeros) and change the order of the sums to obtain
\begin{eqnarray}\label{39}\nonumber
\langle\hat{\sigma}_{x}\rangle&=&\frac{1}{4}\sin(2\theta)e^{-\Gamma t}\sum_{k=0}^{\infty}\Big(\frac{\gamma}{\eta}(1-e^{-2\eta t})\Big)^{k}\langle k| \hat{D}^{\dagger}(\alpha)\hat{\rho}_F(0)\hat{D}(\alpha)  |k\rangle \times\\&& \nonumber\sum_{n=0}^{\infty}\frac{e^{-2\eta n t}}{\Big(\frac{\gamma}{\eta}(1-e^{-2\eta t})\Big)^{n}}\frac{k!}{(k-n)!n!}+c.c.
\end{eqnarray}
The second sum can be cut at $n=k$ as the terms for $n>k$ are zero. therefore it may be summed to give
\begin{eqnarray}
\langle\hat{\sigma}_{x}\rangle&=&\frac{1}{4}\sin(2\theta)e^{-\Gamma t}\sum_{k=0}^{\infty}\Big(\frac{\gamma+i\chi e^{-2\eta t}}{\eta}\Big)^{k}\langle k| \hat{D}^{\dagger}(\alpha)\hat{\rho}_F(0)\hat{D}(\alpha)  |k\rangle +c.c.
\end{eqnarray}
By defining
\begin{eqnarray}
\langle\hat{\sigma}_{x}\rangle&=&\frac{1}{4}\sin(2\theta)e^{-\Gamma t}\sum_{k=0}^{\infty}\Big(\frac{\gamma+i\chi e^{-2\eta t}}{\eta}\Big)^{k}\langle k| \hat{D}^{\dagger}(\alpha)\hat{\rho}_F(0)\hat{D}(\alpha)  |k\rangle +c.c.
\end{eqnarray}
\begin{figure}[h!]
\begin{center}
\includegraphics[scale=0.4]{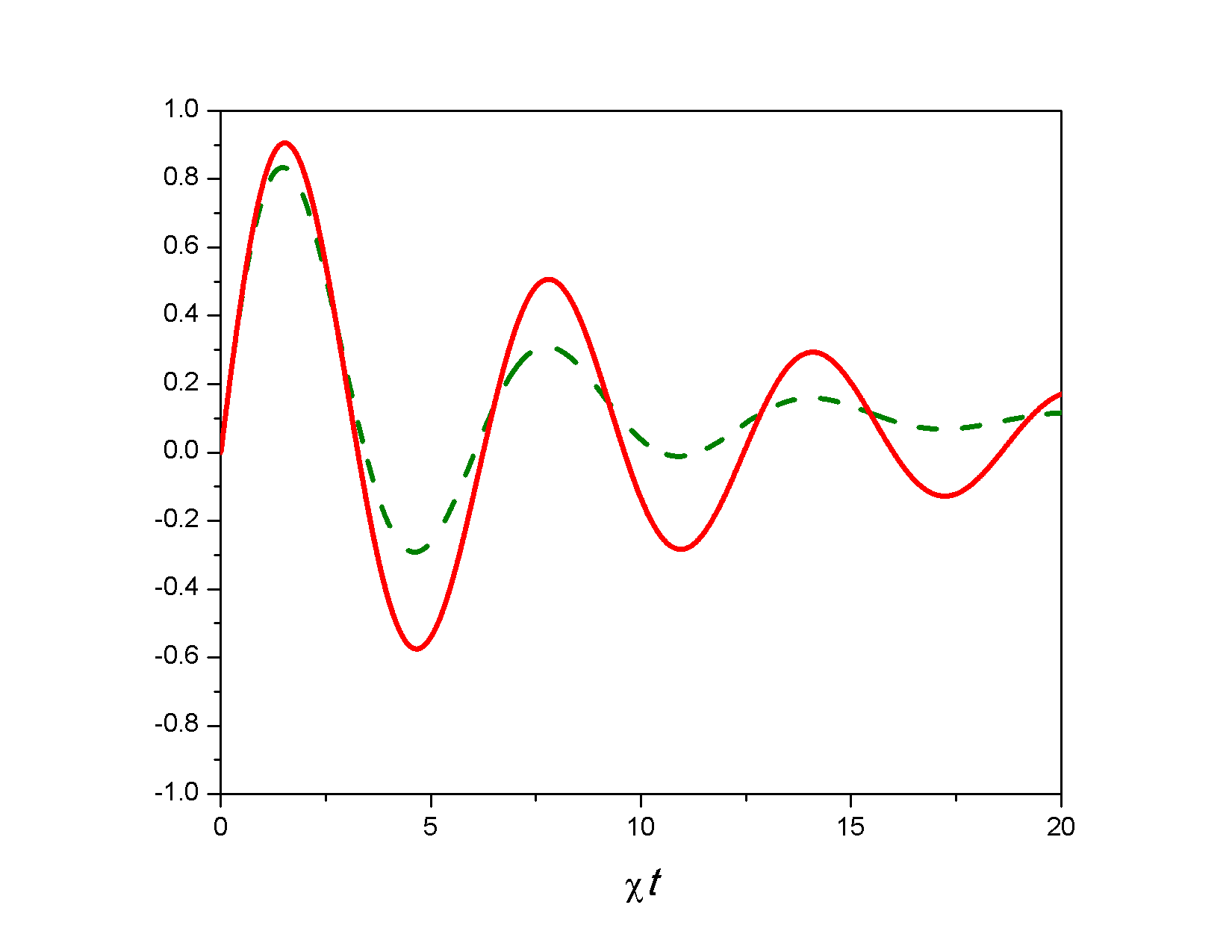}
\caption{Plot of the function  $\epsilon+e^{-2\gamma t}(\sin \chi t-\epsilon \cos \chi t)$ as a function of $\chi t$ for $\gamma=0.05$ and $\epsilon=0.05$ (solid line) and $\gamma=0.1$ and $\epsilon=0.1$ (dashed line).}\label{fig1}
\end{center}
\end{figure}
\begin{eqnarray}\label{40}\nonumber
\mu=\sqrt{\frac{\epsilon^2+e^{-4\gamma t}+2\epsilon\sin{\chi t}e^{-2\gamma t}}{1+\epsilon^2}}, \qquad \tan \phi=-\frac{\epsilon+e^{-2\gamma t}(\sin \chi t-\epsilon \cos \chi t)}{\epsilon^2+e^{-2\gamma t}(\cos \chi t+\epsilon \sin \chi t)},
\end{eqnarray}
with $\epsilon=\gamma/\chi$ we finally obtain

\begin{eqnarray} \label{42}
\langle\hat{\sigma}_{x}\rangle&=&\frac{1}{2}\sin(2\theta)e^{-\Gamma t}\sum_{k=0}^{\infty}\mu^n \cos n\phi\langle k| \hat{D}^{\dagger}(\alpha)\hat{\rho}_F(0)\hat{D}(\alpha)  |k\rangle,
\end{eqnarray}
We plot in figure 1 the numerator of $\tan \phi$, i.e. the function $\epsilon+e^{-2\gamma t}(\sin \chi t-\epsilon \cos \chi t)$ for two values of $\gamma$. Whenever this function crosses zero, $\tan \phi$ is zero. We take in particular the interaction time  of the  first zero, which gives $\phi=\pi$ and define $s=\frac{\mu+1}{\mu-1}$ to write
\begin{eqnarray}
\langle\hat{\sigma}_{x}\rangle&=&\frac{1}{2}\sin(2\theta)e^{-\Gamma t}\sum_{k=0}^{\infty}\left(\frac{s+1}{s-1}\right)^n \langle k|  \hat{D}^{\dagger}(\alpha)\hat{\rho}_F(0)\hat{D}(\alpha)   |k\rangle,
\end{eqnarray}
or finally, comparing to equation (\ref{quasi})
\begin{eqnarray}
\langle\hat{\sigma}_{x}\rangle&=&\frac{(1-s)\pi}{2}\sin(2\theta)e^{-\Gamma t}F(\alpha,s),
\end{eqnarray}
{\it i.e.}, by measuring an atomic observable, namely, the atomic polarization, we can reconstruct a quasiprobability distribution function even though atomic and field decays take place.
\section{Conclusions}
We have solved the dispersive interaction
between a quantized electromagnetic field and a decaying two-level
atom in cavity subject to losses by using superoperator techniques.
We have shown that even in the both decaying cases we can still obtain information
about the initial cavity field by means of $s$-parametrized
quasiprobability distribution functions. Due to the fact that  these functions contain
complete information about the state of the cavity field, we are able to determine the field state, completely.
One thing to consider is the fact that an effective (dispersive) interaction
produces much slower processes such that, both atomic and field decays, may be of importance.

Moreover, if we consider a very small $\theta$, {\it i.e.}, the atom being mostly in the ground state with a very small contribution from the excited state, the reconstruction is still possible, as
\begin{eqnarray}
\langle\hat{\sigma}_{x}\rangle\approx {(1-s)\pi}\theta e^{-\Gamma t}F(\alpha,s),
\end{eqnarray}
and, although the reconstruction would be severely diminished by such a small angle and the fact that we are considering a finite atomic decay rate, it is still possible to obtain whole information from the initial field state.

Finally note that, if we consider $\gamma =0$, {\it i.e.,} an ideal cavity, $\mu=1$ [see equation (\ref{42})] and then the Wigner distribution function may be reconstructed.

\end{document}